\RequirePackage{pdf14}

\documentclass[letterpaper,11pt]{article}
\pdfoutput=1
\usepackage[usenames,dvipsnames,table]{xcolor}
\usepackage{graphicx}
\usepackage{caption}
\usepackage{subcaption}
\usepackage{array,setspace,mathrsfs,amsthm,amsfonts,amsmath,colonequals,mathtools,bbm}
\usepackage{jheppub}
\usepackage{afterpage}
\usepackage{xspace}
\usepackage[normalem]{ulem}
\usepackage{cancel}
\usepackage{slashed}

\newcommand{\p}[1]{(\ref{#1})}

\newcommand{\ep}{\epsilon}
\newcommand{\eps}{\epsilon}

\makeatletter
\def\maketag@@@#1{\hbox{\m@th\normalfont\normalsize#1}}
\makeatother

\newcommand{\beqa}{\begin{eqnarray}}
\newcommand{\eeqa}{\end{eqnarray}}
\newcommand{\beq}{\begin{equation}}
\newcommand{\eeq}{\end{equation}}

\newcommand{\calS}{\mathcal{S}}\newcommand{\calP}{\mathcal{P}}

\newcommand{\abra}[2]{\ensuremath{\langle #1 #2\rangle}}

\newcommand{\HPL}{\text{H}}
\newcommand{\SB}{\text{SB}}

\newcommand{\vac}[1]{\ensuremath{\left< \, #1\, \right>}}

\title{Bootstrapping pentagon functions}

\author{Dmitry Chicherin,}
\author{Johannes Henn,}
\author{Vladimir Mitev}

\affiliation{PRISMA Cluster of Excellence,\\
 Johannes Gutenberg University,\\
 Staudingerweg 9, 55128 Mainz, Germany}

\emailAdd{chicherin@uni-mainz.de}
\emailAdd{henn@uni-mainz.de}
\emailAdd{vmitev@uni-mainz.de}

\preprint{MITP/17-081 }

\bigskip
\abstract{
In {{PRL 116 (2016) no.6, 062001}}, the space of \textit{planar} pentagon functions that describes all two-loop on-shell five-particle
scattering amplitudes was introduced. 
In the present paper we present a natural extension of this space to \textit{non-planar} pentagon functions.
This provides the basis for our pentagon bootstrap program.
We classify the relevant functions up to weight four, which is relevant for two-loop scattering amplitudes.
We constrain the first entry of the symbol of the functions using information on branch cuts.
Drawing on an analogy from the planar case, we introduce a conjectural second-entry condition on the symbol.
We then show that the information on the function space, when complemented with some
additional insights, can be used to efficiently bootstrap 
individual Feynman integrals. The extra information is read off of Mellin-Barnes representations of the integrals,
either by evaluating simple asymptotic limits, or by taking discontinuities in the kinematic variables.
 We use this method to evaluate the symbols of two non-trivial non-planar five-particle integrals, up to and including the finite part.
}


\begin{document}
\setcounter{tocdepth}{2}
\maketitle
\setcounter{page}{1}


\section{Introduction}
\label{sec:introduction}

The idea of bootstrapping scattering amplitudes from their analytic properties 
goes back to the analytic S-matrix program \cite{Eden:1966dnq}. 
When a perturbative expansion is available, for example when perturbative QCD is applicable, the constraints coming for example from perturbative unitarity and from the expected behavior in kinematic limits can be made very concrete. 
At the one-loop level, the relevant function space has been known for a long time. Combining this insight about one-loop Feynman integrals with the above
universal properties turned out to be extremely powerful.
This culminated in the analytic determination
of entires classes of $n$-particle one-loop scattering amplitudes \cite{Bern:1994zx}.

For many years, a big obstacle was the fact that two-loop Feynman integrals 
are far less explored than their one-loop counterparts, and most analytically
known results were for four-particle integrals in various kinematical configurations.

It was realized that a large class of Feynman integrals are represented by iterated integrals, the latter being natural generalizations of the logarithms and dilogarithms appearing in the one-loop case. While these functions had appeared previously in many special cases, the realization that they can be effectively discussed using the so-called symbol \cite{Goncharov:2010jf} 
opened the door to further progress. The symbol encodes the way in which the functions are defined from elementary integrands, the latter being called the alphabet. The symbol and alphabet together encode the analytic structure of the loop integral. 

Further, it was realized that by studying the leading singularities of loop integrals \cite{ArkaniHamed:2010gh} allows one to defined certain \textit{pure integrals}. The latter have a uniform transcendental weight, i.e.~a fixed number of integrations, and involve only numbers as prefactors. In other words, they do not
contain any kinematic prefactors accompanying the transcendental functions.
This concept also turned out to be important for very practical aspects,
as pure functions satisfy canonical differential equations \cite{Henn:2013pwa}. 
From the latter, the properties of the integrals can be read off conveniently.

Once one has an idea of what the symbol alphabet of a certain scattering amplitude is, this is a very powerful constraint on the answer. It reduces the problem to finding  a fixed (albeit sometimes large) number of coefficients. Combining this with other information, for instance about the known structure of amplitudes in soft/collinear or Regge limits, can
sometimes fix the answer completely. 
This modern amplitude bootstrap program was started in \cite{Dixon:2011pw,Dixon:2011nj}, where several  planar six-particle amplitudes in ${\mathcal N} = 4 $ super Yang-Mills were bootstrapped. To date, this program has been pushed to much higher loop orders \cite{Dixon:2014voa,Dixon:2015iva,Caron-Huot:2016owq}, and has also been applied to seven-particle amplitudes \cite{Drummond:2014ffa,Dixon:2016nkn}. 
In most of these cases, the symbol alphabet is conjectured, although supporting evidence comes from explicit calculations of certain loop integrals, and from observed cluster algebra properties of these
scattering amplitudes \cite{Golden:2013xva}. One can also glean insights about the symbol alphabet from
an analysis of the Landau equations of Feynman graphs \cite{Dennen:2015bet}.

So far, this bootstrap program has been limited to planar amplitudes in ${\mathcal N} = 4 $ super Yang-Mills.
The main reason for this is that the latter have a dual conformal symmetry, which considerably simplifies the 
kinematic dependence of the amplitudes. In fact, after taking into account the infrared structure and dual conformal Ward identity, 
the four- and five-particle amplitudes are essentially predicted, so that six- and seven-particle amplitudes are the first interesting cases.

In this paper, we initiate the bootstrap program for the generic QCD case, starting with five-particle amplitudes, both for the planar and non-planar case.\footnote{In principle, one could also attempt to bootstrap for four-particle amplitudes, however the bootstrap becomes more powerful with more external legs, as this allows one to probe various kinematic limits.}
 The most important input into the bootstrap program is the function alphabet.
Our starting point is \cite{Gehrmann:2015bfy}, where all functions relevant for planar two-loop five-particle scattering were computed (see also \cite{Papadopoulos:2015jft, Badger:2016ozq, Badger:2017jhb, Abreu:2017hqn} for related work). 
 The planar function space is described by an alphabet $\mathbb{A}_{\text{P}}$ of $26$ letters. In this paper, we propose that the corresponding non-planar alphabet $\mathbb{A}_{\text{NP}}$ is given by a set of $31$ letters, which is obtained from permutations of the letters of $\mathbb{A}_{\text{P}}$.

We introduce a second generalization of the bootstrap program. While previously the bootstrap was mostly applied  to entire amplitudes, we show how to use it in order to compute individual Feynman integrals. 

One important refinement of the bootstrap program was to incorporate the Steinmann relations, which forbid discontinuities  in overlapping kinematic channels. These relations effectively lead to a constraint on the second entry of the symbol. In the five-particle case, to the best of our knowledge it is not known whether the Steinmann relations imply constraints beyond the Regge limit \cite{Bartels:2008ce}. Here, we make an observation, based on the planar case, that a certain second-entry condition seems to be valid.

When applying the bootstrap method to individual integrals, we can no longer rely on universal properties in kinematic limits. Nonetheless, these limits turn out to be very useful. We extract the information about the limits from Mellin-Barnes representations of the integrals \cite{Gluza:2007rt,Blumlein:2014maa,Dubovyk:2015yba}. 
Although the latter are in general rather complicated, it turns out that they simplify considerably in suitably defined limits. The key point is that it is sufficient to take rather simple limits, reminiscent for example of multi-Regge limits, to fix the parameters in the ansatz. In the limit, the number of Mellin-Barnes integrals is reduced considerably, and the remaining integrals can easily be resummed \cite{Ochman:2015fho}.

We also introduce a further new tool for extracting information from the Mellin-Barnes representations. We show how to compute single and multiple discontinuities of the latter. As the resulting functions have lower transcendental weight, they are much easier to compute.

The outline of the paper is as follows.
In section \ref{sec:alphabet}, we present our conjecture for the non-planar pentagon alphabet, and discuss its properties. 
In section \ref{sec:mb}, we recall how to derive MB representations, and how asymptotic limits and discontinuities can be extracted from them.
In section \ref{sec:application} we apply the above ideas to the calculation of a pair of two-loop non-planar five-point integrals.
We conclude in section \ref{sec: conclusions}.

\section{The non-planar pentagon function alphabet}
\label{sec:alphabet}

\paragraph{Symbols of functions.} Our ultimate goal is to present the results of the pertagon Feynman integrals $I$ in terms of suitable functions such as polylogarithms. Almost as good is to compute the \textit{symbol} $\SB[I]$ of the Feynman integral. We define symbols as follows. Given a set of functions $f_1,\ldots, f_r$ (the alphabet) of the kinematic variables $x=(x_1,\ldots, x_s)$, one defines the weight $n$ symbol $[f_{i_1},\ldots, f_{i_n}]$ iteratively as
\beq
\label{eq: symbol definition}
[f_{i_1},\ldots, f_{i_n}](x)=\int d\log f_{i_n}(x') [f_{i_1},\ldots, f_{i_{n-1}}](x')\,,
\eeq
over some suitable integration path. The symbol does not contain the information of the integration contour or of the values that the iterated integral has to take at the boundary points. These integration constants have to be provided once the symbol is known and it then becomes in principle possible to express the function itself in terms of explicit functions such as polylogarithms. 

Writting a given function $G(x)$ as an iterated integral as in \eqref{eq: symbol definition}, we denote its symbol by $\SB[G(x)]$. We refer to the article \cite{Duhr:2011zq} for a more in-depth introduction on symbols. The advantage of symbols is their ability to capture the main combinatorial and analytical properties of iterated integral functions like polylogarithms, while being significantly easier to deal with. In particular, if the function $G$ is defined via a differential equation, its symbol $\SB[G]$ is in a sense the general solution to the differential equation.

 As an example, let us consider the dilogarithm function $\text{Li}_2(x)$. We have (beware the order inversion)
\beq
\begin{split}
&\text{Li}_2(x)=-\int_{0}^x\frac{\log(1-y)}{y}dy=-\int_{0}^xd\log(x')\int_{0}^{x'}d\log(1-x'')\,,\\
 \Longrightarrow\, &\SB[\text{Li}_2(x)]=-[1-x,x]\,.
 \end{split}
\eeq
We remark that the first entry of the symbol contains information on the its discontinuities (see also section~\ref{subsec: discontinuities}), while derivatives in act on the last entry, for instance $\partial_x \SB[\text{Li}_2(x)]$ is given by $-[1-x]\partial_x \log(x)=-\tfrac{1}{x}[1-x]$. 

For another example of symbols, applying the definition \eqref{eq: symbol definition} to the harmonic polylogs (HPL) of appendix \ref{app: harmonic polylogs}, we find that their symbols are given by
\beq
\label{eq: symbol of the HPL}
\SB[\HPL_{a_1,\ldots, a_n}(x)]=[x-a_n,x-a_{n-1},\ldots, x-a_1]\,.
\eeq

\paragraph{The pentagon alphabet.} The scattering process of five massless particles with momenta $p_i$ is described by five kinematical invariants $v_i$. We introduce the notations: 
\beq
\begin{split}
&v_i = s_{i,i+1}=2p_i\cdot p_{i+1} \,,\qquad 
\textbf{a}_{1,2,3,4} = v_1 v_2 - v_2 v_3 +v_3 v_4 - v_1 v_5 -v_4 v_5\,,\\
&\Delta = \det(2 p_i \cdot p_j)\,.
\end{split}
\eeq
The indices are cyclic, meaning that we set $v_{i+5}\equiv v_i$ and $\textbf{a}_{\cdots (i+5)\cdots }\equiv \textbf{a}_{\cdots i\cdots } $ for all $i$. We remark that  $\textbf{a}_{1,2,3,4} =\text{tr}[\slashed{p}_4\slashed{p}_5\slashed{p}_1\slashed{p}_2]$ and $\Delta=(\text{tr}_5)^2$ with $\text{tr}_5=\text{tr}[\gamma_5\slashed{p}_4\slashed{p}_5\slashed{p}_1\slashed{p}_2]$. We thus also use $\sqrt{\Delta}=\text{tr}_5$. Using the variables $v_i$ has its advantages, though it is often convenient to switch to the $\beta$-variables of \cite{Bern:1993mq}, which have the property that $\sqrt{\Delta}$ can be expressed in them using only rational functions. 

The alphabet $\mathbb{A}_{\text{P}}$ used for the planar five-point amplitudes and integrals was introduced in \cite{Gehrmann:2015bfy}. It is made out of the 26 letters $\mathbb{A}_{\text{P}}=\{W_1,\ldots, W_{20}\}\cup \{W_{26},\ldots, W_{31}\}$ where the \textit{even letters} are (we let the index $i$ run over $1,\ldots, 5$)
\begin{align}
&W_i=v_i=2p_i\cdot p_{i+1}\,,& 
&W_{5+i}=v_{i+2}+v_{i+3}=2p_{i+3}\cdot(p_{i+2}+p_{i-1})\,,&\nonumber\\
&W_{10+i}=v_i-v_{i+3}=2p_{i+2}\cdot(p_{i+3}+p_{i-1})\,,&
&W_{15+i}=v_i+v_{i+1}-v_{i+3}=-2p_i\cdot p_{i+2}\,,&
\end{align}
while the \textit{odd ones} read
\beq
W_{25+i}=\frac{\textbf{a}_{i,i+1,i+2,i+3}-\sqrt{\Delta}}{\textbf{a}_{i,i+1,i+2,i+3}+\sqrt{\Delta}}\,,\qquad \text{ for } i=1,\ldots, 5\,,
\eeq
with the last (even) letter being $W_{31}=\sqrt{\Delta}$. We remark that in the $\beta$ variables of \cite{Bern:1993mq}, all the letters $W_i$ become rational functions.

In this article we call {\it even} the letters $\{W_i\}_{i=1}^{25}$ (and also $W_{31}$) and {\it odd} the letters $\{W_{j}\}_{j=26}^{30}$. When the momenta $p_i$ are real and we use the Minkowski metric, then the complex conjugation is realized as follows: $(\sqrt{\Delta})^{*} = - \sqrt{\Delta}$. Consequently, $(W_j)^* = W_j^{-1}$ for $j = 26, \ldots,30$ and symbols containing an odd number of odd letters change sign.

For the non-planar alphabet $\mathbb{A}_{\text{NP}}$, we need to close $\mathbb{A}_{\text{P}}$ under permutations. The odd letters are closed under permutation but the even ones are not. The minimal option is to simply introduce five additional even letters $\{W_{21},\ldots, W_{25}\}$ given by
\beq
W_{20+i}=v_{2+i}+v_{3+i}-v_{i}-v_{i+1}=2p_{i+2}\cdot(p_i+p_{i+3})\,,\qquad \text{ where } i=1,\ldots, 5\,.
\eeq
Thus, the non-planar alphabet is defined as $\mathbb{A}_{\text{NP}}=\{W_i\}_{i=1}^{31}$.
In the appendix \ref{subsec: permutation group}, we describe the action of the group ${\cal S}_5$ on the alphabet $\mathbb{A}_{\text{NP}}$.

Lastly, we remark that only the 10 letters $\{W_i\}_{i=1}^5\cup\{W_{j}\}_{i=16}^{20}=\{s_{ij}\}_{i<j=1}^5$ appear in the first entries of the non-planar symbols. This is the well-known \textit{first entry condition}, see \cite{Maldacena:2015iua} for an explanation. In the planar case, while all the $\{W_i\}_{i=1}^5\cup\{W_{j}\}_{i=16}^{20}$ remain allowed letters, only the first five $\{W_i\}_{i=1}^5$ are then allowed first entries.

\paragraph{The integrable symbols.} Given the non-planar alphabet $\mathbb{A}_{\text{NP}}$ we are interested in determining \textit{the set of integrable symbols} of given weight that are subject the first entry condition as well as other additional conditions as the case requires. 

We remind that a symbol $S$, written in our alphabet as
\beq
\label{eq: definition S in our alphabet}
S=\sum_{i_1,\ldots,i_n=1}^{31}c_{i_1\cdots i_n}[W_{i_1},\ldots, W_{i_n}]\,,
\eeq
where the $c_{i_1\cdots i_n}$ are constants, is called \textit{integrable} if it satisfies the following integrability condition 
\beq
0\stackrel{!}{=}\sum_{i_1,\ldots,i_n=1}^{31} c_{i_1\cdots i_n}\left\{\frac{\partial \log W_{i_k}}{\partial v_a}\frac{\partial \log W_{i_{k+1}}}{\partial v_b}-(a\leftrightarrow b)\right\}[W_{i_1},\ldots,\hat{W}_{i_k},\hat{W}_{i_{k+1}}, W_{i_n}]\,,
\eeq
for all $k=1,\ldots, n-1$ and all $a,b=1,\ldots, 5$. In the above, $\hat{\cdot}$ indicates omission. The integrability condition guarantees that the iterated integral \eqref{eq: symbol definition} for the symbol $S$ is independent of infinitesimal variations of the integration path. 

The other conditions that we impose on the symbol $S$ are:
\begin{enumerate}
\item  \textit{the first entry condition} which stipulates that in \eqref{eq: definition S in our alphabet} the index $i_1$ only runs over the set $\{1,\ldots, 5\}$ for $\mathbb{A}_{\text{P}}$ and over $\{1,\ldots, 5\}\cup\{16,\ldots, 20\}$ for $\mathbb{A}_{\text{NP}}$.
\item \textit{the second entry condition}. This condition is more hypothetical, which is why we also performed the integrable symbol classification without it. It corresponds to forbidding the appearance in \eqref{eq: definition S in our alphabet} of the terms $[W_1,W_8,\cdots]$, $[W_5,W_8,\cdots]$ and their permutations. Such terms could in principle appear in the planar integrable symbols, but happen to not appear in planar Feynman integrals. We conjecture that they are also absent from the non-planar Feynman integrals. Explicitly, the forbidden pairs of indices are
\beq
\footnotesize{
\begin{split}
&\{1, 8\}\,, \{1, 9\}\,, \{1, 14\}\,, \{1, 15\}\,, \{1, 24\}\,, \{1, 25\}\,, \{2, 9\}\,, \{2, 
  10\}\,, \{2, 11\}\,, \{2, 15\}\,,\{2, 21\}\,, \{2, 25\}\,, 
  \\ 
  &\{3, 6\}\,, \{3, 10\}\,, \{3, 
  11\}\,, \{3, 12\}\,, \{3, 21\}\,, \{3, 22\}\,, \{4, 6\}\,, \{4, 7\}\,,\{4, 12\}\,, \{4, 
  13\}\,, \{4, 22\}\,, \{4, 23\}\,,\\ 
  & \{5, 7\}\,, \{5, 8\}\,, \{5, 13\}\,, \{5, 14\}\,, \{5, 
  23\}\,, \{5, 24\}\,, \{16, 8\}\,, \{16, 10\}\,, \{16, 11\}\,, \{16, 14\}\,,
   \{16, 21\}\,, \\&\{16, 
  24\}\,, \{17, 6\}\,, \{17, 9\}\,, \{17, 12\}\,, \{17, 15\}\,, \{17, 22\}\,, \{17, 25\}\,, \{18, 
  7\}\,,\{18, 10\}\,, \{18, 11\}\,, \\&  \{18, 13\}\,, \{18, 21\}\,, \{18, 23\}\,, \{19, 6\}\,, \{19, 
  8\}\,, \{19, 12\}\,, \{19, 14\}\,, \{19, 22\}\,, \{19, 24\}\,, \{20, 7\}\,,\\& \{20, 9\}\,, \{20, 
  13\}\,, \{20, 15\}\,, \{20, 23\}\,, \{20, 25\}\,.
\end{split}}
\eeq
\end{enumerate}
The results of the classification of the integrable symbols in the alphabets $\mathbb{A}_{\text{P}}$ and $\mathbb{A}_{\text{NP}}$ are presented in Table~\ref{tab:number of integrable symbols}.

\begin{table}[h]
\centering
\renewcommand{\arraystretch}{1.6}
\begin{tabular}{|l|c|c|c|c|}
\hline
Weight
& 
1
&
2
&
3
&
4
\\
\hline
$\#$ of integrable symbols for $\mathbb{A}_{\text{P}}$ & 5$\,|\,$0 & 25$\,|\,$0 & 125$\,|\,$1 & 635$\,|\,$16\\
after 2nd entry condition & 5$\,|\,$0 & 20$\,|\,$0 & 80$\,|\,$1 & 335$\,|\,$11\\
$\#$ of integrable symbols for $\mathbb{A}_{\text{NP}}$ & 10$\,|\,$0 & 100$\,|\,$9 
& 1000$\,|\,$180 & 9946$\,|\,$2730\\
after 2nd entry condition & 10$\,|\,$0 & 70$\,|\,$9 & 505$\,|\,$111 & 3736$\,|\,$1191\\
\hline
\end{tabular}
\renewcommand{\arraystretch}{1.0}
\caption{ We list here the number of independent integrable symbols of given weight for the planar and the non-planar alphabets. In each case, we indicate the number of even$\,|\,$odd symbols.}
\label{tab:number of integrable symbols}
\end{table}

\section{Mellin-Barnes Technology}
\label{sec:mb}

In this section, we introduce several tools involving Mellin-Barnes integrals that we will then make use of in section~\ref{sec:application} in order to compute an explicit Feynman integral. 

\subsection{Mellin-Barnes representations for non-planar Feynman integrals}

Deriving Mellin-Barnes (MB) representations for Feynman integrals is a very standard procedure, see e.g. \cite{Gluza:2007rt,Smirnov:2012gma}.
One starts from a Feynman parametrized form of the answer, factorizes the integrand with the help of the basic Mellin-Barnes integral
formula, 
\begin{align}\label{MBone}
\frac{1}{(X+Y)^a} = \frac{1}{\Gamma(a)} \int_{c-i \infty}^{c+i\infty} \frac{dz}{2\pi i} \Gamma(-z) \Gamma(a+z) X^z Y^{-a-z} \,,
\end{align}
where the $z$-integration goes along the vertical axis with real part $c\in(-a,0)$
and then finally carries out the Feynman parameter integrals. 

The resulting Mellin-Barnes representation is not unique and for example the number of integration parameters can depend on the way the representation is introduced. For example, in the planar case, it is often advisable to proceed loop-by-loop, to obtain a low-dimensional representation.

In the non-planar case, special care has to be taken to obtain an MB representation with good convergence properties. 
The issue is that factors such as $(-1)^z$ can lead to bad convergence properties in the complex plane \cite{Czakon:2005rk}, and it
is better to avoid them. While special tricks may work for individual cases, in general it seems best to start with the global
Feynman parametrization, as opposed to the loop-by-loop approach \cite{Smirnov:2012gma,Dubovyk:2015yba}.
Moreover, ref. \cite{Dubovyk:2015yba} suggests that one can make a suitable choice of the projective delta function in the Feynman parametrization
in order to arrive at a relatively low-dimensional MB integral.

Another new feature of the non-planar case is the appearance of new kinematic invariants.
For example, in the four-point case, one obtains a MB representation of the type \cite{Tausk:1999vh}
\begin{align}\label{exampleMB4point}
\int dz_{i} (-s)^{z_1} (-t)^{z_{2}} (-u)^{z_{3}} g(z_i) \,.
\end{align}
Here $p_{i}^2=0,\, \sum_{i=1}^{4} p_i = 0,\, s= 2 p_1 \cdot p_2 ,\, t=2 p_2 \cdot p_3$ and $u=2 p_1 \cdot p_3$.
Since $s+t+u=0$, at least one of the factors in the integrand leads to a negative power of $-1$. 
For this reason, it is better to start the calculation in a more general kinematic regime, where $s,t,u$ are considered independent,
and only at the end of the calculation impose $s+t+u=0$ \cite{Tausk:1999vh}.
As we will see later, similar features occur for our non-planar pentagon integrals.

For example, consider the Feynman diagram of topology (c) of table 3 of \cite{Bern:2015ple}, see also figure~\ref{Fig:topology c}.
Deriving the global Feynman representation, it is clear that one can find a MB representation
of the type (\ref{exampleMB4point}), with exponentials of the following factors: 
\begin{align}\label{factorstopoc}
\{ -s_{12}, -s_{23}, -s_{34}, -s_{45}, -s_{15}, -s_{35}, -s_{14} \}\,.
\end{align}
Similarly to the four-point example, only five of these variables are independent. 
In the present case, one can however directly use an independent set of variables,
while not having any exponentials of $(-1)$ in the MB integrand.
This can be seen as follows. If we choose $\{ s_{34}, s_{45}, s_{15}, s_{14}, s_{35} \}$ as independent, then we have 
$s_{12} = s_{34} + s_{35} + s_{45}\,, s_{23} = s_{14} + s_{15} + s_{45}$.
Therefore, we can consider a kinematic region where all factors in eq. (\ref{factorstopoc}) are positive.
This point will also be important when considering kinematic limits and discontinuities.

\subsection{Suitable kinematic limits}
\label{subsec: suitable kinematic limits}

Thanks to our bootstrap hypothesis,
we do not need to compute the complicated MB integral, 
but rather it is sufficient to extract some information from it.
To this end we can define kinematic limits that considerably simplify
the integrals.

In the case of scattering amplitudes, a useful limit is the multi-Regge limit,
where the kinematic variables are rescaled according to
\begin{align}
\label{limitmultiregge}
s_{12} \to t_1 \,,\quad  s_{23} \to t_2 \,,\quad  s_{34} \to s_2/\rho \,,\quad  s_{45} \to s_1 /\rho \,,\quad s_{15} \to s/\rho^2 \,,
\end{align}
with $\rho \to 0$.
Analyzing the limit at the level of the symbol, we see that the alphabet $\mathbb{A}_{\text{NP}}$
simplifies to
\begin{align}
\{ \rho , s, s_1, s_2 ,s_1 + s_2,  t_1,t_2,t_1 + t_2 \}
\end{align}
It is obvious that this alphabet decomposes into smaller independent alphabets.
This implies that the result will be given by products of simpler functions.
The only non-trivial type of function, corresponding to the  is the $3$-letter alphabet $\{ s_1, s_2 ,s_1+ s_2\}$ (and similar for $s \leftrightarrow t$)
gives rise to harmonic polylogarithms \cite{Remiddi:1999ew}. The latter are single-variable functions that can usually be obtained in a relatively simple way by transforming the MB integrals into sums (see section \ref{sec:resummingMB}).

Similarly, we can take limits reminiscent of soft limits, by setting $\rho \to 1/\rho$ before taking the limit. Moreover, we can consider permutations of those limits, and in this way obtain additional information.

Considering that we are not using any information from the expected properties of {\it {amplitudes}} in the limit, but rely on the MB representation to extract that information, we are free to consider other limits as well. For example, in the case of integral topology (i) of table 3 of \cite{Bern:2015ple}, discussed in section~\ref{sec:application}, it turns out that one can find a nice MB representation depending on the independent variables $\{ s_{12} , s_{13}, s_{23}, s_{24}, s_{34} \}$. In this case one can consider a limit analogous to (\ref{limitmultiregge}), but where the $s_{i,i+1}$ are replaced by those variables, see section~\ref{eq: limits of MB integrals}.

Since we only need to determine a limited number of information, we can choose the limits that are most accessible. Of course, further limits can be used as valuable cross checks, thereby giving additional support to the bootstrap hypothesis. We find that in most cases it is sufficient to compute the limits to leading power of the expansion, matching the logarithmically enhanced terms $\log^k(\rho)$ and the finite part in the $\rho \to 0$ limit with the ansatz. In principle, one can also consider power suppressed terms to provide additional information (see section \ref{sec:application} for an example of this).

\subsection{Resumming Mellin-Barnes integrals}
\label{sec:resummingMB}

One dimensional Mellin-Barnes integrals of only one scale can often be evaluated explicitly in terms of harmonic polylogarithms (HPL) and rational functions. The procedure works as follows. We use the package {\tt{MBsums}} provided by \cite{Ochman:2015fho}. The contour is closed in such a way as to have the resulting sum be a Laurent expansion around $x=0$. This means that if the scale $x$ enters the MB integral over $z$ as $x^z$, then we have to close the contour to the right and if it appears as $x^{-z}$, then we close to the left. 
As an example, consider one of the typical MB integrals that appear
\beq
I_{\text{example}}=\int_{-\frac{1}{2}-i\infty}^{-\frac{1}{2}+i\infty}\frac{dz}{2\pi i}x^{z} \frac{\Gamma (1-z) \Gamma (-z) \Gamma (z+1) \psi(-z-1)}{\Gamma (-z-1)}\,,
\eeq
where $\psi$ is the digamma function. Using {\tt{MBsums}} and closing the contour to the right, we obtain the sum representation
\beq
I_{\text{example}}=1+\sum_{k=2}^\infty\frac{(-1)^{k} k! }{(k-2)!}x^{k-1}\left(h_{k-2}-2 h_{k}-\log (x)+\gamma_{\rm E} \right)\,,
\eeq
where $h_k=\sum_{\ell=1}^k\frac{1}{\ell}$ are the harmonic numbers. We can now very easily expand $I_{\text{example}}$ to arbitrary order around $x=0$. By matching terms with an ansatz of the type $p_i(x) \text{HPL}(A,x)$, we can evaluate the sum.\footnote{We would be remiss not to mention the powerful package \textsf{Xsummer}, see \cite{Moch:2005uc} or \textsf{Sigma}, see \url{https://www.risc.jku.at/research/combinat/software/Sigma/} and \cite{CarstenSchneider}, that can also help one perform the sum. For our purposes, they were not needed. } The prefactors $p_i(x)$ are of the type 
\beq
p_i(x)\in \mathbb{P}_{n,m}\equiv \left\{\frac{P_n(x)}{Q_m(x)}\, :\, \begin{array}{l}P_n\text{ is a polynomial of degree $n$ and }\\Q_m=x^a(x-1)^b(x+1)^c \text{ with }\max(a,b,c)\leq m \end{array}\right\}\,.
\eeq
We increase $n$, $m$ and the HPL weight $A$ until we find a solution. We observe that a linear basis of the space of prefactors $\mathbb{P}_{n,m}$ is given by 
\beq
\mathbb{P}_{n,m}=\text{span}\left(1,x,\ldots, x^n, x^{-1},\ldots, x^{-m},\frac{1}{x-1},\ldots \frac{1}{(x-1)^m},\frac{1}{x+1},\ldots \frac{1}{(x+1)^m}\right)\,.
\eeq
For the example at hand, we obtain finally the result
\beq
I_{\text{example}}=\frac{1}{(1+x)^3}\left(1 + \gamma_{\rm E}-3x-x^2+2 x\,\HPL_{-1}(x)-2 x\,\HPL_0(x)\right)\,,
\eeq
For the evaluation and series expansion of the harmonic polylogarithms, we use the package {\tt HPL} of \cite{Maitre:2005uu}. 
The methods described in this section can be generalized to multiple MB integrals of a single scale, though in that case a better approach than using {\tt MBsums} is to use the package\footnote{See \url{https://mbtools.hepforge.org/}.} {\tt MBasymptotics}.

\subsection{Discontinuities}
\label{subsec: discontinuities}

Consider a function $f(x)$ that is real-valued for $x>0$, and that may have a branch cut starting somewhere along
the negative real axis. Then we define the discontinuity according to
\begin{align}
\label{eq: definition of the discontinuity}
{\rm Disc}_{x} f(x)_{x=-y} := \frac{1}{2\pi i} \left[ f( y e^{-i \pi}) - f(y e^{i \pi}) \right]  \,,\; y>0\,. 
\end{align}
Given a MB representation $f(x) = \int dz \,x^z g(z)$, this
yields
\begin{align}
{\rm Disc}_{x} f(x)_{x=-y} = -  \int dz\, y^z \frac{g(z)}{\Gamma(-z)\Gamma(1+z)} \,.
\end{align}
By definition, the r.h.s. is only defined for $y>0$. Let us see how this works in a few examples.
First, we have the identity 
$-\frac{1}{1-x} \log x = \int \frac{dz}{2 \pi i} x^z \Gamma^2(-z) \Gamma^2(1+z)$.
Taking a discontinuity, we obtain
\begin{align}
{\rm Disc}\left(-\frac{1}{1-x} \log x\right)_{x=-y}=-\frac{1}{1+y} = -  \int \frac{dz}{2 \pi i} y^z \Gamma(-z) \Gamma(1+z) \,.
\end{align}
For the second example, we want to distinguish the cases where the branch cut starts from $0$ from those where it starts somewhere else on the negative real axis. This is important to distinguish symbol terms such as e.g. $[x,  ...]$ and $[1+x,...]$. Consider for instance
\begin{align}
g(x) = -\frac{1}{1+x} \log(1+x) = \int\frac{dz}{2\pi i} x^z \Gamma(-z)\Gamma(1+z) \left( \psi(1+z) + \gamma_{\rm E}\right) \,,
\end{align}
where we remind that $\psi$ is the digamma function.
Then we have
\begin{align}
h(y) = {\rm Disc}_{x} g(x) |_{x=-y} = - \int \frac{dz}{2\pi i} y^z  \left( \psi(1+z) + \gamma_{\rm E}\right) \,,
\end{align}
We can verify that we can rewrite this function as $h(y) = \frac{\theta(y-1)}{y-1} .$
In fact, if $0<y<1$, the contour can be closed on the right, leading to a vanishing result.
If $y>1$, the contour can only be closed on the left, and this leads to the expected result.

\section{Explicit integrals from Mellin-Barnes representations}
\label{sec:application}

In this section we want to provide explicit applications of the symbol classification of table~\ref{tab:number of integrable symbols} by computing two different two-loop Feynman integral. Specifically, using the notation for pentagon integrals of \cite{Bern:2015ple}, we consider an integral of topology (i), see figure~\ref{Fig:topology i} and one of topology (c), see figure~\ref{Fig:topology c}.

\subsection{Mellin-Barnes representations for the topology (i)} 
\label{secMB6} 

\begin{figure}[htbp!]
             \begin{center}       
              \includegraphics[width=6cm]{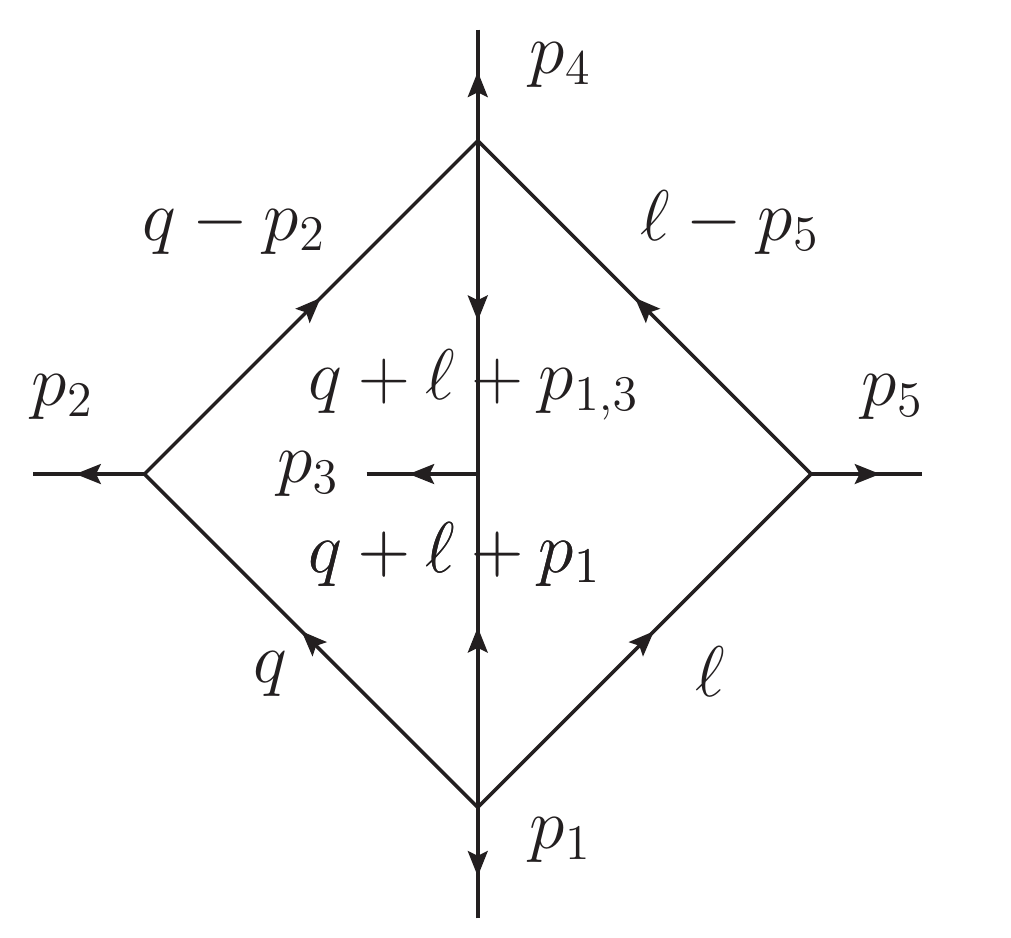}
              \caption{The 2-loop non-planar integral corresponding to topology (i) of \cite{Bern:2015ple}.}
              \label{Fig:topology i}
            \end{center}
\end{figure}

First, we want to derive Mellin-Barnes (MB) representations for the integral of topology (i) of figure~\ref{Fig:topology i} in two ways.

\paragraph{The six kinematic invariants case:} We introduce an auxiliary parameter $\alpha$ to combine a pair of massless propagators with $p^2= 0$. A direct integration shows that
\begin{align} 
\label{WL}
\frac{1}{\ell^2 (\ell + p)^2} = \int_0^1 \frac{d \alpha}{(\ell + \alpha p)^4}\,.
\end{align} 
Then we use this formula to rewrite the diagram of figure~\ref{Fig:topology i} as ($D = 4 - 2\ep$)
\begin{align}
I_{(i)}=\int d^{D} \ell\, d^D q \int \frac{d \alpha_1 \, d \alpha_2 \, d \alpha_3 }{(\ell - \alpha_1 p_5)^4 (q - \alpha_2 p_2)^4 (q+\ell + p_1 + \alpha_3 p_3)^4}\,,
\end{align} 
where the integrations $\alpha_1$, $\alpha_2$, $\alpha_3$ are over $[0,1]^3$.
Then we straightforwardly integrate out the loop momenta $q$ and $\ell$ 
and obtain the result
\begin{align}
I_{(i)}=\int d\alpha_1d\alpha_2d\alpha_3\,\pi^D \frac{\Gamma^3(-\ep)\Gamma(2+2\ep)}{\Gamma(-3\ep)} F_{(i),1}^{-2-2\ep}\,,
\end{align}
where we have defined 
\begin{align}
F_{(i),1}=(-s_{12}) \alpha_2 + (-s_{13}) \alpha_3 + (-s_{15}) \alpha_1 + \alpha_1 \alpha_2 (-s_{25}) + \alpha_1 \alpha_3 (-s_{35}) + \alpha_2 \alpha_3 (-s_{23})\,.
\end{align}
We can now introduce a fivefold MB representation for $F_{(i),1}$ and integrate out the auxiliary parameters $\alpha_1,\alpha_2,\alpha_3$. In this way, we find the following MB representation for $I_{(i)}$
\beq
\label{MB6}
\begin{split}
I_{(i)}=& \pi^D \frac{\Gamma^3(-\ep)}{\Gamma(-3\ep)} \int \frac{[d z]}{(2\pi i)^5} \prod_{j=1}^5 \Gamma(-z_j)
\Gamma(2 + 2\ep +z_{1,2,3,4,5})  \\ 
& \times \frac{\Gamma(-1-2\ep-z_{1,2,4})\Gamma(-1-2\ep-z_{1,3,5}) \Gamma(1+z_{1,4,5})}{\Gamma(-2\ep-z_{1,2,4})\Gamma(-2\ep-z_{1,3,5}) \Gamma(2+z_{1,4,5})}\\&\times  (-s_{15})^{z_1} \, (-s_{12})^{z_2} \, (-s_{13})^{z_3} \, (-s_{25})^{z_4} \, (-s_{35})^{z_5} \, (-s_{23})^{-2-2\ep-z_{1,2,3,4,5}} \,,
\end{split}
\eeq
where we have used the shorthand $z_{i_1,\cdots,i_n}=z_{i_1}+\cdots+z_{i_n}$ and $[dz] = d z_1 \ldots d z_5$.
This MB integral depends on the following six kinematic invariants, not all of which are independent:
\beq
\begin{split}
 \label{kin6}
&s_{13} = v_4 - v_1 -v_2\,,\quad s_{25} = v_3 - v_1 - v_5\,,\quad s_{35} = v_1 -v_3 - v_4\,,\\
&s_{12} = v_1\,, \quad s_{23} = v_2\,\quad s_{15} = v_5\,.
\end{split}
\eeq

\paragraph{The five kinematic invariants approach:} Again, we first use eq. \p{WL} for two pairs of propagators, namely those of the left and middle of the diagram.  In this way we find 
\begin{align}
I_{(i)}=\pi^{\frac{D}{2}} \frac{\Gamma^2(-\ep)\Gamma(2+\ep)}{\Gamma(-2\ep)} \int^1_0 d \alpha_2 \int^1_0 d \alpha_3 \int d^D \ell 
\frac{1}{\ell^2 (\ell-p_5)^2 (\ell + p_1 + \alpha_2 p_2 + \alpha_3 p_3)^{4+2\ep}}
\end{align}
which is a triangle diagram with massless propagators (integrated over auxiliary parameters $\alpha_2, \alpha_3$) 
and the external momenta 
$p_5$, $p_1 +\alpha_2 p_2 + \alpha_3 p_3$, and $p_4 + \bar\alpha_2 p_2 + \bar\alpha_3 p_3$.
Then we introduce Feynman parameters $x_1, x_2, x_3$ in a standard way for the triangular diagram and find after some manipulations the five-fold MB integral representation
\beq
\label{eq: five fold representation for Ii}
\begin{split}
I_{(i)}=&\, \pi^D \frac{\Gamma^3(-\ep)}{\Gamma(-2\ep)\Gamma(-3\ep)} \int \frac{[d z]}{(2\pi i)^5}\prod_{j=1}^5 \Gamma(-z_j) \Gamma(1+z_{1,2})\Gamma(1+z_{2,3}) \Gamma(1+z_{1,2,3})  \\ 
&\times\, \Gamma(2 + 2\ep +z_{1,2,3,4,5}) \Gamma(-1-2\ep-z_{1,2,3})\Gamma(-1-2\ep-z_{1,2,3,4})\\&\times\,\Gamma(-1-2\ep-z_{1,2,3,5})
\Gamma^{-1}(-2\ep-z_{1,4})\Gamma^{-1}(-2\ep-z_{3,5}) \, (-s_{12})^{z_1} \, (-s_{13})^{z_3} \\
&\times\, (-s_{24})^{z_4} \, (-s_{34})^{z_5} \, (-s_{23})^{-2-2\ep-z_{1,3,4,5}}
\end{split}
\eeq
This MB depends on five kinematic invariants which can be taken as independent variables.

\subsection{The symbol for the topology (i)}
\label{sec: symbol for the topology i}

\paragraph{Classification:} The leading singularity analysis of \cite{Bern:2015ple} finds just one leading singularity of the integral $I_{(i)}$, namely $\tfrac{1}{\sqrt{\Delta}}$. This suggests that the diagram of figure~\ref{Fig:topology i} can be written as  
\beq
\label{eq: top i ansatz}
I_{(i)}=\frac{1}{\sqrt{\Delta}} \sum_{i \geq 0} \frac{1}{\ep^{2-i}} {\cal P}_{i+2}\,,
\eeq
where ${\cal P}_{i}$ is a pure function of weight $i$. Since $\sqrt{\Delta}$ is odd under complex conjugation and $I_{(i)}$ is even, the functions $P_{i}$ have to be odd. We work at symbol level, hence we need the classification of symbols, specifically of the \textit{odd symbols} of weight $\geq 2$, see table~\ref{tab:number of integrable symbols}. We read off 9 integrable symbols at weight 2, 180 at weight 3 and 2730 at weight 4. The diagram obviously has only the following seven nontrivial two-particle cuts 
\beq
\label{eq: first entries for top i}
s_{12} = v_1 \;, \; s_{34} = v_3 \; , \; s_{45} = v_4 \; , \; s_{15} = v_5 \; , \; s_{13} = v_4 - v_1 - v_2 \; , \; 
s_{24} = v_5 - v_2 -v_3\,.
\eeq
Thus, only these kinematic invariants can appear in the first entries of the symbols. Using this, we find 1 integrable symbol at weight 2; 13 at weight 3; 143 at weight 4. We further notice that the diagram has a discrete ${\cal S}_{2} \times {\cal S}_{3}$ permutation symmetry due to the permutations of external points $(1,4)$ and $(2,3,5)$. Imposing this condition, we find 1 integrable symbol at weight 2; 4 at weight 3; 21 at weight 4. Finally, imposing the second entry condition, we find 1 integrable symbol at weight 2; 3 at weight 3; 12 at weight 4. In the following we will work with the symbol ansatz. We summarize the number of symbols in table~\ref{tab:number of symbols for topology i}.

\begin{table}[h]
\centering
\renewcommand{\arraystretch}{1.6}
\begin{tabular}{|l|c|c|c|}
\hline
Weight
&
2
&
3
&
4
\\
\hline
$\#$ of odd symbols for the topology (i) & 9 & 180 & 2730\\
only first entries of \eqref{eq: first entries for top i} & 1 & 13 & 143\\
${\cal S}_{2} \times {\cal S}_{3}$ symmetry & 1 & 4 & 21\\
second entry condition & 1 & 3 & 12\\
\hline
\end{tabular}
\renewcommand{\arraystretch}{1.0}
\caption{ We list here the number of independent symbols of given weight that can contribute to the integral of topology (i).}
\label{tab:number of symbols for topology i}
\end{table}

\paragraph{Computing the symbol:} Having established the ansatz \eqref{eq: top i ansatz} for $I_{(i)}$, where the ${\cal P}_{i}$ are linear combinations of the symbols of the penultimate row of table~\ref{tab:number of symbols for topology i}, we can now start calculating the coefficients of these linear combinations.
To compute the symbol of $I_{(i)}$, we use the MB representation \p{MB6} and then take discontinuities (single and double) of the MB representation and compare them with the discontinuities of the symbol ansatz \eqref{eq: top i ansatz}. We do this for each term in $\epsilon$, so we expand \p{MB6}
\beq
\label{eq: MB topology i each epsilon coefficient}
I_{(i)}=\frac{I_{(i),2}}{\epsilon^2}+\frac{I_{(i),3}}{\epsilon}+I_{(i),4}+\mathcal{O}(\epsilon)\,,
\eeq
where each term $I_{(i),n}$ is given by a MB integral. This is easily done by using the Mathematica packages {\tt MB.m} of \cite{Czakon:2005rk} and {\tt MBasymptotics.m}, see \url{https://mbtools.hepforge.org/}.

We will mostly deal with discontinuities of the MB integrals with respect to the kinematic invariants $s$ at $s \sim 0$ which appear in the MB integrand in the form $(-s)^z$. Thus, we can expand the MB integral at $s \sim 0$ (that usually lowers the dimensionality of MB integrations) and pick up the $\log(-s)$ terms.  
While not all discontinuities have this form, already these easily accessible discontinuities provide many constraints. 

At weight 4 however, this above approach shows its limits. In order to fix the weight 4 of the ansatz \eqref{eq: top i ansatz}, we will also need to consider the discontinuities in $s$ at finite $s$. Furthermore, taking double discontinuities of weight four MB integrals, we obtain functions of maximal weight 2. The corresponding MB integral can be rather easily evaluated using Cauchy's theorem and series summation formulae.

\subsubsection{The \texorpdfstring{$1/\ep^2$}{1/epsilon^2} piece}

At weight 2, we need to fix only one coefficient in the ansatz, i.e. the normalization coefficient $c_2$. Explicitly, our ansatz for the ${\cal P}_2$-piece of \eqref{eq: top i ansatz} is
\beq
\begin{split} 
\label{w2ans}
\SB[{\cal P}_2] =& c_2\Big(-[W_1,W_{30}] - [W_3,W_{26}] + [W_4,W_{26}] + [W_4,W_{30}] + [W_5,W_{26}] \\&+ [W_5,W_{30}] - [W_{16},W_{26}] - [W_{17},W_{30}]\Big)\,.
\end{split} 
\eeq
We compute first a discontinuity in $s_{12}$ at $s_{12} =v_1 \to 0$. To take the discontinuity in $v_1$ of the symbol we just replace $[W_i,*] \to \delta_{i,1} [*]$ 
where we remind that $W_1 = v_1$. In the remaining expression we take $v_1 \to 0$. Applying this operation to the symbol ansatz \p{w2ans}
we find 
\begin{align} 
\label{S2s12}
\mbox{Disc}_{v_{1} \sim 0}\, \SB[{\cal P}_{2}] =  -c_2[W_{30}] = c_2\big([v_2-v_4] + [v_3] - [v_4] - [v_5]\big)\,.  
\end{align}
In the limit $v_1 \to 0$ we have $\lim_{v_1\to 0}\sqrt{\Delta}= v_2 v_3 -v_3 v_4 +v_4 v_5$.

The MB integral \p{MB6} depends on six two-particle invariants, see \p{kin6}, and they cannot all be negative at $v_1 \to 0$. However we can forget for a moment about momentum conservation and take discontinuities of the MB integral in $s_{12}$ at $s_{12} \sim 0$. To do it we just expand \eqref{kin6} at $s_{12} \sim 0$ and pick up the $\log s_{12}$ terms. The result is a two-fold MB integral which involves only $s_{13}, s_{15}, s_{23}, s_{25}$. Now we reintroduce the momentum conservation and in the limit $v_1 \to 0$ we find the kinematic variables
\begin{align}
\label{discs12kin}
s_{13} = v_4 -v_2\,,\quad  s_{25} = v_3 - v_5\,,\quad  s_{23} = v_2\,,\quad  s_{15} = v_5\,,
\end{align}
which all can be negative. This is what we want to avoid $(-1)^z$ issues with MB integrations.

Then we can easily take double discontinuity in any of these four variables (when it is also small): $s_{13} \sim 0$, $s_{25} \sim 0$, $s_{23}  \sim 0$, or $s_{15} \sim 0$. For instance, to take the discontinuity in $s_{15}$ at $s_{15} \to 0$, we expand the two-fold MB integral at $s_{15} \sim 0$ and pick up the $\log(-s_{15})$ term which comes out to be a rational function. Explicity, we find
\begin{align}     
\mbox{Disc}_{v_5 \sim 0} \mbox{Disc}_{v_1 \sim 0} \, I_{(i),2} = \frac{-3}{v_3(v_2-v_4)}\stackrel{!}=\mbox{Disc}_{v_5 \sim 0} \mbox{Disc}_{v_1 \sim 0} \, \frac{\SB[{\cal P}_2]}{\sqrt{\Delta}} \stackrel{\p{S2s12}}{=} -\frac{c_2}{\sqrt{\Delta}}\big|_{\substack{v_5\sim 0\\ v_1\sim 0}}\,.
\end{align}
Recalling that $\sqrt{\Delta} = v_3(v_2-v_4)$ at $v_1,v_5 \to 0$, we
immediately fix the normalization coefficient $c_2=3$.\footnote{A comment is due on the choice of branch for $\sqrt{\Delta}$. The functions $\calP_i$ (and hence $c_2$) change sign under the choice of branch but as long as one is consistent, the final result for the integral $I_{(i)}$ is invariant.}

We can check our result by computing a number of other discontinuities. 
For instance, we find 
\beq
\begin{split}     
&\mbox{Disc}_{v_2 \sim 0} \mbox{Disc}_{v_1 \sim 0} \, I_{(i),2} = \mbox{Disc}_{s_{25} \sim 0} \mbox{Disc}_{v_1 \sim 0} \, I_{(i),2}=  0 \,,\\
&\mbox{Disc}_{s_{13} \sim 0} \mbox{Disc}_{v_1 \sim 0} \, I_{(i),2} = \frac{3}{v_2 v_5}\,,
\end{split}
\eeq
which is obviously compatible with ansatz discontinuity \p{S2s12}.
Let us note that we can easily find discontinuities in $v_3$ or $v_4$ of the symbol \p{S2s12} but not of the MB integral $\mbox{Disc}_{s_{12}\sim 0}I_{(i),2} $. This is due to the fact that the latter depends explicitly on the set \p{discs12kin}, hence its MB integrand lacks a $(-v_3)^z$ or $(-v_4)^z$ factor.

The symbolic expression is a crucial step towards a functional representation for an integral of uniform transcendentality. Once it is known, one can find the 
``beyond-the-symbol" terms (or ``boundary terms" in the language of the differential equations method).
For the symbol $\SB[{\cal P}_2]$ of \eqref{w2ans} with $c_2=3$, we find a particularly simple functional expression  
\begin{align} 
\label{funep2}
{\cal P}_2 = 6\left[ \mbox{Li}_2(W_{26}) + \mbox{Li}_2(W_{30}) - \mbox{Li}_2(W_{26} W_{30}) - \frac{1}{2}\log W_{26} \log W_{30} - \frac{\pi^2}{6} \right] \,.
\end{align} 
This result has also been cross checked numerically.
Let us note that the function ${\cal P}_2$ and its ${\cal S}_5$ permutations span the 9-dimensional odd subspace of weight 2 of the
non-planar pentagon functions, see table~\ref{tab:number of integrable symbols}. Of course, acting with ${\cal S}_5$ permutations on \p{funep2} we obtain many more different functions, hence they have to satisfy some identities. All of them are simple dilogarithm identities, except for the 15-term identity which we present in appendix \ref{AppOddFun}.

\subsubsection{The \texorpdfstring{$1/\ep$}{1/epsilon} piece}

We know from table~\ref{tab:number of symbols for topology i} that we need to fix 4 coefficients in the odd weight 3 symbol ansatz $\SB[{\cal P}_3]$. Similarly to the case for the $1/\epsilon^2$ symbol, we consider first a discontinuity in $s_{12}$ at $s_{12} \to 0$ of the MB integral $I_{(i),3}$ of \eqref{eq: MB topology i each epsilon coefficient}. This yields a two-fold MB integral involving only $s_{13}, s_{15}, s_{23}, s_{25}$ (recall \p{discs12kin})
as well as a  $\log(-s_{12})$ factor. We see that all these two-particle invariants can be chosen negative. We then take a discontinuity in $v_{5}$ of the MB integral at $v_5 \sim 0$. After expanding the two-fold MB integral at $s_{15} \to 0$ and picking up the $\log(-s_{15})$ terms, the 
MB integrations disappear and we find 
\begin{align} \label{discv5discv1}    
\mbox{Disc}_{v_5 \sim 0} \mbox{Disc}_{v_1 \sim 0} \, I_{(i),3} = 6\,\frac{\log(-v_1) - \log(-v_3) + \log(-v_5)}{v_3(v_2-v_4)}\,.
\end{align}
This is already sufficient to fix $\SB[{\cal P}_3]$ completely. 
The other discontinuities that we compute are
\begin{align} \label{}    
\mbox{Disc}_{s_{13} \sim 0} \mbox{Disc}_{v_1 \sim 0} \, I_{(i),3} = 6\,\frac{\log(-v_2) - \log(-v_1) + \log(v_2-v_4)}{v_2 v_5}\,,
\end{align}
as well as 
\begin{align} \label{discv5discv1}    
\mbox{Disc}_{v_1 \sim 0} \mbox{Disc}_{v_1 \sim 0} \, I_{(i),3} =6 \,\frac{\log(-v_4) + \log(-v_5) - \log(-v_3) - \log(v_2 - v_4)}{v_2 v_3 - v_3 v_4 + v_4 v_5}\,.
\end{align}
The explicit result for the symbol $\SB[{\cal P}_3]$ (as well as $\SB[{\cal P}_2]$ and $\SB[{\cal P}_4]$) is provided in an auxiliary file. The precise function expression of ${\cal P}_3$ is not yet known. A convenient way of representing the full function is in terms of iterated integrals, in the spirit of \cite{Caron-Huot:2014lda}. This can be done by specifying a boundary point in the iterated integral, and by matching the boundary value against the limits presented here.

\subsubsection{The \texorpdfstring{$\ep^0$}{epsilon^0} piece}

If we ignore the second entry condition, we need to fix the 21 coefficients in the odd weight four symbol ansatz $\SB[{\cal P}_4]$.

\paragraph{Discontinuities in $s_{12}$ at $s_{12} \to 0$.} Taking a discontinuity of $I_{(i),4}$ in $s_{12}$ at $s_{12} \sim 0$ yields a two-fold MB integral involving only $s_{13}, s_{15}, s_{23}, s_{25}$ which are \p{discs12kin}: $s_{13} = v_4 -v_2$, $s_{25} = v_3 - v_5$, $s_{23} = v_2$ and $s_{15} = v_5$ as well as the $\log(- s_{12})$, $\log^2 (-s_{12})$ factors. We can then take a discontinuity in $v_{5}$ of $v_5 \sim 0$. After expanding the resulting twofold MB integrals at $s_{15} \to 0$ and picking up the $\log (-s_{15})$ terms, the MB integrations disappear and we find 
\begin{align} \label{w3discv5v1}    
\mbox{Disc}_{v_5 \sim 0} \mbox{Disc}_{v_1 \sim 0} \,I_{(i),4} = -6\,\frac{(\log(-v_1) - \log(-v_3) + \log(-v_5))^2}{v_3(v_2-v_4)}\,.
\end{align}
In the above we have ignored terms proportional to $\pi$ since we are working with symbols. 
Comparing \p{w3discv5v1} with the ansatz for $\SB[{\cal P}_4]$, we fix 17 coefficients. 

Then we can consider a discontinuity in $v_1$, since there are $\log(-s_{12})$ factors in the MB integral. This discontinuity is given by a two-fold MB integral. Evaluating it by summing up the residues, we find a series which can be summed up explicitly to give
\beq
\label{w3discv1v1}
\begin{split}
\mbox{Disc}_{v_1 \sim 0} \mbox{Disc}_{v_1 \sim 0} \, I_{(i),4} = &\frac{6}{\sqrt{\Delta}}\biggl( 2 \mbox{Li}_2(1-v_4/v_2) + 2 \mbox{Li}_2(1-v_3/v_5) \\ 
&+ \log^2(1-v_4/v_2) + \log^2(-v_1) \log \frac{v_3}{v_5} \Bigl(1- \frac{v_2}{v_4}\Bigr) \biggr)\,,
\end{split}
\eeq
where $\Delta = v_2 v_3 - v_3 v_4 + v_4 v_5$ in the limit $v_1 \to 0$ and we have dropped any terms proportional to $\pi$. 
Comparing \p{w3discv1v1} with the ansatz we fix one more coefficient. 

We still need to fix two more coefficients. $\mbox{Disc}_{v_2 \sim 0} \mbox{Disc}_{v_1 \sim 0}$ as well as others easily accessible discontinuities do give any further constraints, so we need to consider discontinuities in $s$ at finite values of $s$.

\paragraph{Discontinuities at finite values of the kinematic variables.} Before taking double discontinuities, we consider the asymptotics of $J_{(i),4}=\mbox{Disc}_{v_1 \sim 0} I_{(i),4}$ at $v_5 \to 0$. This results in 0-fold and 1-fold MB integrals.
The 1-fold MB integrals involves nontrivially only $s_{13}$ and $s_{23}$, i.e. $(s_{13}/s_{23})^z$. In the limit $v_1,v_5 \to 0$ we have $s_{13} = v_4 -v_2$ 
and $s_{23} = v_2$. So we can safely throw away the $\log(-s_{12}) , \log(-s_{15}), \log(-s_{25})$ factors contained in the MB integral for they could not mix with the MB integrations. Now we want to calculate $\mbox{Disc}_{s_{13}}J_{(i),4}$ at finite $s_{13}$ as well as $\mbox{Disc}_{s_{23}}J_{(i),4}$ at finite $s_{23}$. We define the discontinuity as in \eqref{eq: definition of the discontinuity} and evaluate the integrals using residues.  
We have to be cautious closing the integration contour when we take a discontinuity in $s_{13}$ since $-s_{13} \ll -s_{23}$. Thus for an integral containing $(s_{13}/s_{23})^z$ we close the contour on the right.
Let us stress that the choice of the contour is important, closing the contour in the opposite way we would obtain wrong results. 
Thus we obtain
\beq
\mbox{Disc}_{s_{13}} \mbox{Disc}_{v_1 \sim 0} \, I_{(i),4} = 6\, \frac{(\log(v_2-v_4) - \log(-v_2))^2}{v_3(v_2-v_4)}\,,\qquad 
\mbox{Disc}_{s_{23}} \mbox{Disc}_{v_1 \sim 0} \, I_{(i),4} = 0\,, 
\eeq   
up to terms proportional to $\pi$.
Comparing the first of the previous equations with the ansatz we find one more constraint.

In a very similar fashion, we obtain
\beq
\mbox{Disc}_{v_{5}} \mbox{Disc}_{v_1 \sim 0} \,  I_{(i),4} = -6\, \frac{(\log(v_5-v_3) - \log(-v_5))^2}{v_4(v_5-v_3)}\,,\qquad 
\mbox{Disc}_{s_{25}} \mbox{Disc}_{v_1 \sim 0} \,  I_{(i),4} = 0  \,,
\eeq
again up to terms proportional to $\pi$. Comparing the first of the previous equations with the ansatz for $\SB[{\cal P}_4]$ we fix the last two coefficients.

We remind that the explicit result for the symbol $\SB[{\cal P}_4]$ is provided in an auxiliary file. Furthermore, we remark that the symbol of $I_{(i)}$ has also been computed using the differential equation method in \cite{AJTprivate}. 

\subsection{Limits of the Mellin-Barnes integrals}
\label{eq: limits of MB integrals}

We can provide independent cross checks on the computation of the integral $I_{(i)}$ by taking kinematic limits, as explained in section \ref{subsec: suitable kinematic limits}. For example, we can consider the soft-like limit
\begin{align}
\label{eq: example limit of MB integral for the variables}
   s_{12} = t_1 \,,\quad  s_{13} = \rho s_{2}\,,\quad  s_{23} = \rho s_{1}\,,\quad  s_{24} = \rho^2 s\,,\quad  s_{34} = t_{2} \,,
\end{align}
with $\rho \to 0$.
Starting from the five-fold MB representation  \eqref{eq: five fold representation for Ii} for the integral $I_{(i)}$, we find that at most one-fold MB representations survive after taking the limit.

The remaining integrals are rather simple and can be evaluated analytically as described in section \ref{sec:resummingMB}.
We types of integrals we encounter are
\beq
\begin{split}
&\int_{c-i\infty}^{c+i\infty}  x^{-z}  \Gamma(-1 - z)^2 \Gamma(1 + z)^2  = -x \Big[\HPL_{2,0}(x) + \HPL_{0^3}(x)+2\zeta_2  \HPL_0(x)\Big] \,,  \\
&\int_{c-i\infty}^{c+i\infty}  x^{-z}  \Gamma(-1 - z)^2 \Gamma(1 + z)^2 \psi(-1-z)
      =  x \Big[\left(\zeta_3+2\gamma_{\rm E} \zeta_2\right) \HPL_0(x)\\&\qquad \qquad + \zeta_2 \left(\HPL_{0^2}(x)- \HPL_2(x)\right)+ \gamma_{\rm E} \left( \HPL_{2,0}(x)+ \HPL_{0^3}(x)\right)- \HPL_{2,1,0}(x)+ \HPL_{0^4}(x)\Big]
 \,.  
\end{split}
\eeq
with $x=t_{1}/t_{2}$ and $-1<\text{Re}(c)<0$. These integrals are needed at weight $3$ and $4$, respectively.  In this way, we obtain
 \beq
 \begin{split}
 \lim_{\rho\rightarrow 0}\SB[I_{(i),2}]\,=\,&
 12 [\rho, \rho] + 3 [\rho, s] + 6 [\rho, s_{2}] - 6 [\rho, t_{1}] - 
 3 [\rho, t_{2}] + 3 [s, \rho] + 3 [s, s_{2}] \\& - 3 [s, t_{1}] 
 +6 [s_{2}, \rho] + 3 [s_{2}, s] - 3 [s_{2}, t_{2}] - 6 [t_{1}, \rho] - 
 3 [t_{1}, s]\\& + 3 [t_{1}, t_{2}] - 3 [t_{2}, \rho] - 3 [t_{2}, s_{2}] + 
 3 [t_{2}, t_{1}]\,.
 \end{split}
 \eeq
 We omit from writing out the $1/\eps$ and $\eps^0$ terms of $\SB[I_{(i)}] $ solely for reasons of space. This limit alone determines the all coefficients in our ansatz at $1/\eps$ and $\eps^0$.  At $\eps^0$, it fixes $10/12$ coefficients if one uses the second entry condition. The remaining coefficients are fixed by the discontinuity calculation of section \ref{sec: symbol for the topology i}.  Alternatively, we could have also considered slightly more complicated limits to fix them.

\subsection{The topology (c) integral with two magic numerators}

\begin{figure}[htbp!]
             \begin{center}       
              \includegraphics[width=6cm]{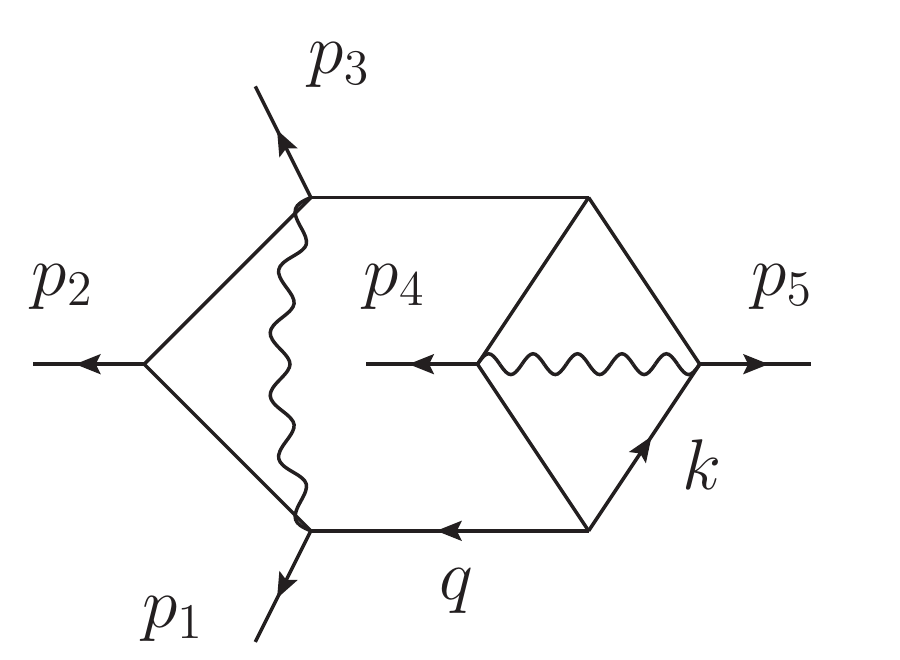}
              \caption{The 2-loop non-planar integral $I_{(c)}$ with numerator \eqref{eq: numerators of Ic}.}
              \label{Fig:topology c}
            \end{center}
\end{figure}

Having computed $I_{(i)}$, let us turn our attention to a more complicated case, namely that of the 4D non-planar integral $I_{(c)}$ of figure~\ref{Fig:topology c}. It is a \textit{finite} integral in four dimensions, since its numerator is a product of two ``magic" numerators (see \cite{ArkaniHamed:2010gh})
\beq
\label{eq: numerators of Ic}
\text{numerator of $I_{(c)}$}\, =\, \vac{1|q(q-p_1-p_2)|3} \vac{4|(k+q)k|5}\,,
\eeq
and it has a \textit{single leading singularity}. 
We want to introduce a Feynman parametrization for this integral and we start with its box subdiagram. Since there are no divergences, we only need its finite part. According to \cite{Dixon:2011ng}, it coincides with the 6D box integral (without numerator) times a $\abra{4}{5}$ factor carrying the helicity charge. Then we introduce a parametric representation for the 6D two-mass-easy box and obtain
\begin{align}
\begin{array}{c}
\includegraphics[width = 2.8cm]{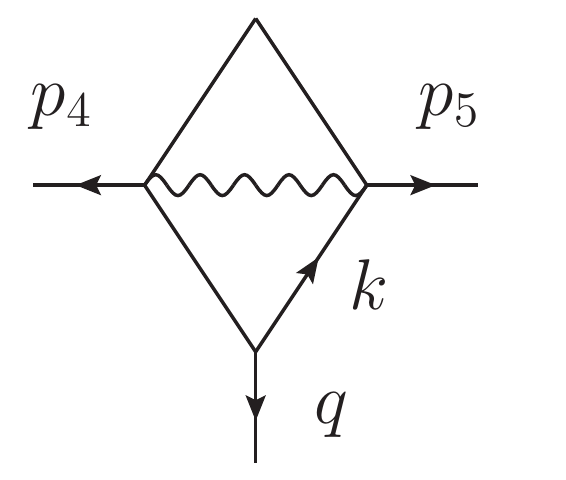}
\end{array} = \abra{4}{5} \int^1_0 d \alpha \int^1_0 d \beta \frac{1}{(q+\alpha p_4 + \beta p_5)^2} \,.
\end{align}
We insert this subgraph in the 4D diagram of figure~\ref{Fig:topology c} and we obtain the 4D pentagon with a magic numerator (three legs are massless and two legs are massive). Using \cite{Drummond:2010mb}, we find
\beq
\begin{split}
\begin{array}{c}
\includegraphics[width = 4cm]{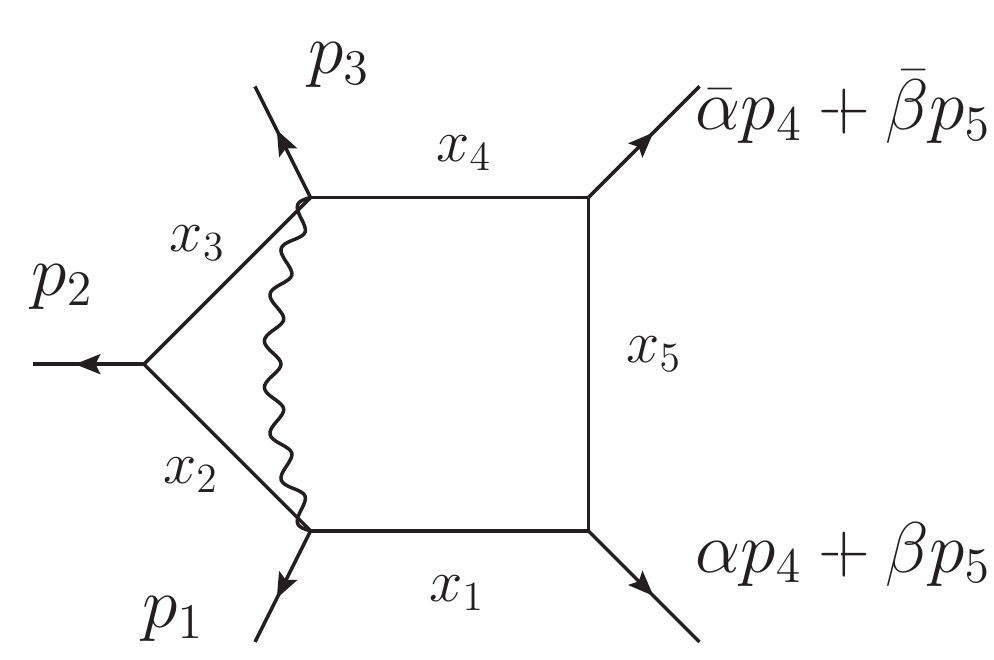}
\end{array} = \int [d x]  \frac{x_5}{F_{(c)}^3} [45]\Big[ \abra{1}{4}\abra{5}{3} \alpha\bar\beta -  \abra{1}{5}\abra{4}{3} \bar\alpha\beta\Big]\,, 
\end{split}
\eeq
where we have defined $\bar{\alpha}\equiv 1-\alpha$, $\bar{\beta}\equiv 1-\beta$, the $x_1 , \ldots , x_5$ are Feynman parameters with $[dx] = \delta(\sum_i x_i-1) \prod_i d x_i$ and 
\beq
\label{eq: Feynman polynomial for the topology c}
\begin{split}
F_{(c)}\, =\,& x_1 x_3 s_{12} + x_1 x_4 s_{45} + x_2 x_4 s_{23} + x_1 x_5 \alpha\beta s_{45} + x_2 x_5 (\alpha\beta s_{45} + \alpha s_{14} + \beta s_{15})\\
 &+ x_3 x_5 (\bar\alpha\bar\beta s_{45} + \bar\alpha s_{34} + \bar\beta s_{35}) + x_4 x_5 \bar\alpha\bar\beta s_{45} \,.
\end{split}
\eeq
Thus, we finally obtain the Feynman parametrization for our diagram:
\beq
\label{FPtopc}
I_{(c)} = \int^1_0 d \alpha \int^1_0 d \beta  \int [d x]  \frac{x_5}{F_{(c)}^3} s_{45}\left( \abra{1}{4}\abra{5}{3} \alpha\bar\beta -  \abra{1}{5}\abra{4}{3} \bar\alpha\beta\right)\,.
\eeq
We know that the leading singularity of $I_{(c)}$ is $\frac{1}{[13][45]}$.\footnote{This follows from the fact that the leading singularity of the box sub-diagram is $\frac{1}{[45]}$. Inserting this leading singularity in the whole diagram we obtain again the box with magic numerator whose leading singularity is $\frac{1}{[13]}$.} So we should have in total
\begin{align}
I_{(c)}= \frac{1}{[13][45]} \, f_{(c)}\,, 
\end{align}
where $f_{(c)}$ is a \textit{pure} function of weight four. In order to study the discrete symmetries of $f_{(c)}$ we rewrite equation \p{FPtopc} in the following form:
\beq
f_{(c)} = -s_{45}\int^1_0 d \alpha \int^1_0 d \beta  \int [d x]  \frac{x_5}{F_{(c)}^3} \left( \abra{1}{4}[45]\abra{5}{3}[31] \alpha\bar\beta + \abra{1}{5}[54]\abra{4}{3}[31] \bar\alpha\beta\right)\,.
\eeq
Let us define a pair of $\mathbb{Z}_2$ transformations that exchange the external momenta: $\sigma: p_4 \leftrightarrows p_5$ and $\tilde\sigma: p_1 \leftrightarrows p_3$. Due to the numerator \eqref{eq: numerators of Ic}, the full integral $I_{(c)}$ changes sign when acted upon with $\sigma$ or $\tilde{\sigma}$. 
Furthermore, these transformations act of the F-polynomial as
$\sigma(F_{(c)}) = F_{(c)}\big|_{\alpha \leftrightarrows \beta}$ and $\tilde\sigma(F_{(c)}) = F_{(c)}\big|_{\alpha \leftrightarrows \bar\alpha, \beta \leftrightarrows \bar\beta, x_1 \leftrightarrows x_4, x_2 \leftrightarrows x_3 }$. 
Consequently,  we observe that $f_{(c)}$ is $\mathbb{Z}_2 \times \mathbb{Z}_2$ symmetric, i.e. $\sigma(f_{(c)} ) = \tilde\sigma(f_{(c)} ) = f_{(c)}$.
Independently of the this $\mathbb{Z}_2\times \mathbb{Z}_2$ symmetry, the pure function $f_{(c)}$ can be decomposed in parity even and odd pieces as $f_{(c)}=  f_{(c)}^{\rm even} + f_{(c)}^{\rm odd}$. Introducing the integrals
\begin{align} 
\label{IaIb}
J_a = s_{45}\int^1_0 d \alpha \int^1_0 d \beta  \int [d x]  \alpha\bar\beta \frac{x_5}{F_{(c)}^3} \,,\qquad 
J_b = s_{45}\int^1_0 d \alpha \int^1_0 d \beta  \int [d x]  \bar\alpha\beta \frac{x_5}{F_{(c)}^3}\,,
\end{align}
that are related to each other as $\sigma(I_a) = I_b$, we obtain for the parity odd/even pieces\footnote{We remind of the relation $\abra{i}{j}[jl]\abra{l}{m}[mi]=\tfrac{1}{2}\big(s_{ij}s_{lm}-s_{il}s_{jm}+s_{im}s_{jl}-4i\varepsilon(i,j,l,m)\big)$, where $\varepsilon(i,j,l,m)=\varepsilon_{\mu\nu\rho\sigma}p_i^{\mu}p_j^{\nu}p_l^{\rho}p_m^{\sigma}$ is completely antisymmetric. Due to our sign conventions of resolving the square root we have $4i\varepsilon(1,4,5,3)=\sqrt{\Delta}$.}
\beq
\begin{split}
f_{(c)}^{\rm odd} \,&=\,\frac{\sqrt{\Delta}}{2}(J_a-J_b)\,,\\ f_{(c)}^{\rm even} \,&=\,\frac{1}{2}\Big[(s_{14}s_{35} - s_{15} s_{34} +s_{13} s_{45})J_a+(s_{15}s_{34} - s_{14} s_{35} +s_{13} s_{45})J_b\Big]\,.
\end{split}
\eeq
Since the F-polynomial $F_{(c)}$ contains $11$ terms, we expect to get a  10-fold MB representation for $J_{a/b}$. However, introducing the MB integral in a naive way we find zero due to a factor of $1/\Gamma(0) = 0$. To avoid this, we use an analytic regularization in \p{IaIb}, substituting $F_{(c)}^{-3} \to F_{(c)}^{-3-\ep}$. Following this,  we introduce an MB representation and after resolving the singularities in $\ep$, we obtain a 6-fold MB integral. However, it turns out to be more useful to derive a MB representation that has only five different kinematic invariants. Simply expressing everything in \eqref{eq: Feynman polynomial for the topology c} in terms of $s_{14}, s_{15}, s_{34}, s_{35}$ and $s_{45}$, we obtain then the following 10-fold MB representation
\begin{align}
J_a\,=&\,\int \frac{[d z]}{(2\pi i)^{10}}\frac{(-s_{14})^{-1+z_7-z_{8,9,10}} (-s_{34})^{-1-z_{2,3,4,5}} (-s_{45})^{z_{3,4,8,9}-z_{6,7}}(-s_{35})^{z_{2,5}}(-s_{15})^{z_{6,10}}}{\Gamma (1-z_{5,10}) \Gamma (3+z_{1,2,3,4,5,8,9,10})}\notag\\
&\times \prod_{j=1}^{10}\Gamma(-z_j) \Gamma (1-z_{10})  \Gamma (-z_5) \Gamma (-z_{1,2,3}) \Gamma (1+z_{8,9,10})^2 \Gamma (z_{6,7}-z_9) \\
&\times \Gamma (1+z_{1,2,3,9}) \Gamma (1+z_{1,2,3,4,5}) \Gamma (2+z_{1,2,3,4,5}) \,,\notag
\end{align}
with a similar expression for $J_b$.

\subsection{Computing the symbol of topology (c)}

Having computed in the previous section the MB representations of the functions $f_{(c)}^{\text{odd/even}}$, we now want to compute their symbols $\SB[f_{(c)}^{\text{odd/even}}]$. We begin by imposing the first entry condition specific to the integral $I_{(c)}$, followed by the second entry condition and finally the $\calS_2 \times \calS_2$ discrete symmetries of the maps $\sigma$ and $\tilde{\sigma}$. In table~\ref{tab: symbols for topology c} we show the number of integrable symbols remaining after each condition is applied. The last line contains the number of integrable symbols that enters in the next part of the computation.

\begin{table}[h]
\centering
\renewcommand{\arraystretch}{1.6}
\begin{tabular}{|l|c|c|}
\hline

& 
Weight 4 odd
&
Weight 4 even
\\
\hline
$\#$ of integrable symbols &  2730 & 9946 \\
after the first entry condition for $I_{(c)}$ &  220& 2435\\
after the second entry condition &106 &  970 \\
$\calS_2 \times \calS_2$ symmetry & 23 & 272 \\
\hline
\end{tabular}
\renewcommand{\arraystretch}{1.0}
\caption{ We list here the number of independent integrable symbols that contribute to the integral of topology $(c)$. }
\label{tab: symbols for topology c}
\end{table}

In general, our strategy can be summarized as follows. We want to take sufficiently simplifying limits of the MB integral representations of $f_{(c)}^{\text{odd/even}}$ and compare with the same limit of the symbol ansatz from the last row of table~\ref{tab: symbols for topology c}. In order to make the limits calculable, we want to take some of the $s_{ij}$ to either 0 or 1 in order to simplify the alphabet and bring it either to the HPL alphabet $\{1+x,x,1-x\}$ (see appendix \ref{app: harmonic polylogs} and \cite{Maitre:2005uu}) or to a 2dHPL alphabet\footnote{This nice set of functions can be obtained by using the alphabet $\{x,1+x,y,1+y,x+y,1+x+y\}$,  classifying the integrable symbols up to weight 3 and obtaining their functional realization using just the $\log$, $\text{Li}_2$ and $\text{Li}_3$ functions. For our purposes here, weight 3 is enough, because we can take derivatives of the weight 4 symbols in order to catch subleading terms. } depending on two variables $x$ and $y$, see \cite{Gehrmann:2000zt}. This allows us to easily convert symbols to functions, ignoring irrelevant boundary terms proportional to $\zeta$-values. We know how to obtain  the asymptotic expansions of these functions, which means that we can also obtain the asymptotic expansions of the symbol ansatz. This allows us to compare the ansatz with the limit of the MB integral. In particular, we find many homogeneous constraints by comparing log-pieces of the asymptotics, i.e.~logs that are present in the expansion of the ansatz but absent in the MB integral.

\paragraph{Soft and Regge limits.} Specifically, a set of homogeneous equations for the integral $I_{(c)}$ can be obtained by considering soft-like
and Regge-like limits of the kinematic invariants as in section~\ref{subsec: suitable kinematic limits}. Specifically, like in \eqref{limitmultiregge}, we perform replacements like $s_{14} \to t_1$, $s_{15} \to t_2$, $s_{34} \to s_2/\rho$, $s_{35} \to s_1 /\rho$, $s_{45} \to s/\rho^2$ (as well as all possible 120 permutations thereof) and then consider the limit $\rho\to 0$ (soft) or $\rho\to \infty$ (Regge). In such limits, the MB integrals simplify, but are still too involved to compute easily. Nevertheless, by just looking at which kinematic invariants and which powers of $\log(\rho)$ are still presents in the limit of the MB integral, we can obtain many constraining equations for the symbols $\SB[f_{(c)}^{\text{odd}}]$.

\paragraph{Inhomogeneous constraints.} 
For the odd function $f_{(c)}^{\text{odd}}$, the homogeneous constraints obtained by the above approach are sufficient to fix all coefficients up to a normalization, which can then be computed using an inhomogeneous equation. An inhomogeneous constraint can for example be obtained by taking the MB representation, setting $s_{45}= t u^{\frac{1}{2}}$  and $s_{14}= t u^{-\frac{1}{2}}$ and then taking in sequence the limits $s_{15}\to 0$, $s_{34}\to 0$ and $t\to 0$. The resulting 2-fold MB integral can be expanded in a power series in $u$ using {\tt{MBasymptotics}}. Reconstructing the power series into HPL functions as in section~\ref{sec:resummingMB}, we find in that limit for both the odd and the even function
\beq
\label{eq: inhom limit of f}
\begin{split}
\lim f_{(c)}^{\text{even/odd}}\,=\,&-\big[\HPL_{-3}(u)- \HPL_{-2, -1}(u) - \zeta_2\HPL_{-1}(u)\big] \times \Big[\HPL_0(s_{15}) -  \HPL_0(t)+\frac{1}{2}\HPL_0(u)\Big]\\&+\zeta_2 \HPL_{-1}(u) \big(\HPL_0(s_{34}) - \HPL_0(s_{35})\big) + W^{\text{even/odd}}_4(u)\,,
\end{split}
\eeq
where $W^{\text{even/odd}}_4(u)$ are weight four functions that depend only on $u$ and have no logarithmic singularities at $u\rightarrow 0$. This condition is then sufficient to fix $\SB[f_{(c)}^{\text{odd}}]$

 To finish the calculation of the symbold of the even function $f_{(c)}^{\text{even}}$, we have to do a bit more and compute other limits similar to \eqref{eq: inhom limit of f}. The computations become harder and we have to consider limits of the MB integral in which we can only obtain the first few terms in a power series expansion. One such limit is attained by first taking $s_{14}\to 0$, then $s_{15}\to 0$, followed by setting $s_{34}=s_{35}=\sqrt{x}$ and $s_{45}=\tfrac{1}{\sqrt{x}}$. After taking that limit, we can compute the first few terms in the power series expansion using {\tt{MBasymptotics}} and the {\tt{PSLQ}} algorithm:
\beq
\begin{split}
\lim f_{(c)}^{\text{even}}\,=&\,x \left(\log ^2(x)-6 \log (x)+12\right)+\frac{1}{4} x^2 \left(7 \log ^2(x)-2 \log (x)-23\right)\\&+\frac{1}{36} x^3 \left(-110 \log ^2(x)-139 \log (x)+372\right)+\mathcal{O}\left(x^4\right)\,,
\end{split}
\eeq
up to factors proportional to $\zeta_2$, $\zeta_3$ and $\zeta_4$. Comparing the above expansion with corresponding limit of the symbol ansatz allows us to fix many coefficients. 
Using many such laborious steps, we can fix the symbol $\SB[f_{(c)}^{\text{even}}]$ exactly. The answer is provided in an ancillary file.

\section{Conclusions and Outlook}
\label{sec: conclusions}

In this article, we presented a conjecture for the alphabet $\mathbb{A}_{\text{NP}}$ of non-planar pentagon functions. We expect that these functions describe on-shell five-particle scattering amplitudes at two loops, and possibly also at higher loop orders. 

Based on experience with the planar case, we presented a conjectural \textit{second-entry condition} for the symbol alphabet. The latter considerably reduces the space of allowed functions. It would be interesting if this condition could be proven, perhaps by some variant of the Steinmann relations.

As a first application, we bootstrapped the symbols $\SB[I_{(i)}]$ and $\SB[I_{(c)}]$ of two non-planar two-loop integrals. 
These integrals are needed for the computation of the non-planar scattering amplitudes in ${\mathcal N}=4$ sYM. They depend in an intricate way on the five-particle kinematics. 
The fact that the symbols are consistent with all discontinuities and limits considered is a strong consistency check that the answer is correct. For the integral $I_{(i)}$, we also independently verified the solution by comparing against results from the differential equation approach \cite{AJTprivate}. We anticipate that the method can also be effectively used to bootstrap integrals with more propagator factors. It would be interesting to push this method of calculating Feynman integrals further and to compute the remaining topologies of \cite{Bern:2015ple}. 

A natural further avenue of future research is to bootstrap full amplitudes starting from the function space. Possible applications include non-planar Yang-Mills or (super)gravity theories. Doing so requires having a good control over the space of rational functions multiplying the symbols in the amplitude. One can in principle obtain these rational functions by computing the leading singularities of the corresponding Feynman diagrams.

\section*{Acknowledgments}
The are indebted to Adriano Lo Presti and Thomas Gehrmann for providing a complementary validation of $I_{(i)}$ \cite{AJTprivate} using the differential equation approach. We are also very thankful to Maximilian Stahlhofen and to Pascal Wasser for many useful discussions.
Furthermore, we thank the authors of \cite{Dubovyk:2015yba} for useful correspondence regarding Mellin-Barnes representations for non-planar Feynman integrals.
J.M.H. thanks KITP for hospitality in 2016 and 2017 during the programs ``LHC Run II and the Precision Frontier'', ``Scattering Amplitudes and Beyond''s and the MIAPP program ``Mathematics and Physics of Scattering Amplitudes'', where part of this work was done.
The authors are supported in part by the PRISMA Cluster of Excellence at Mainz university.
This project has received funding from the European Research Council (ERC) under the European Union's Horizon 2020 research and innovation programme (grant agreement No 725110), ``Novel structures in scattering amplitudes''.


\appendix

\section{Appendix}
\label{sec:appendix}

\subsection{The permutation group}
\label{subsec: permutation group}

The permutation group ${\cal S}_5$ has 7 irreducible representations labeled by Young diagrams (YD). We also label them by their dimension
\begin{itemize}
\item The trivial representation $1$ of YD $[5]$.
\item The sign representation  $1'$ of YD $[1,1,1,1,1]=[1^5]$. 
\item The standard representation $4$ of YD $[4,1]$.
\item The standard representation tensored with the sign representation of YD $[2,1,1,1]=[2,1^3]$. We label it by $4'$. 
\item The five dimensional representation $5$ of YD $[3,2]$.
\item The five dimensional one $5'$ of YD  $[2,2,1]$, given by the tensor product of the $5$ with the $1'$. 
\item The $6$ (given by the exterior tensor product of $[4,1]$ with itself) with YD $[3,1,1]$. 
\end{itemize}
The characters tables of ${\cal S}_5$ are well-known and it is hence a rather simple exercise to compute the projection matrices and the decomposition of a given representation if one knows how the transpositions $P_{i,i+1}$ for $i=1,2,3,4$, act.

It is also useful to consider the action of the permutation group ${\cal S}_5$ on the external momenta  $p_1,\ldots,p_5$, which then implies that the space of 31 letters $\mathbb{A}_{\text{NP}}$ of section \ref{sec:alphabet} decomposes into representation of ${\cal S}_5$. The action of ${\cal S}_5$ on the letters $W_i$ is non-linear but it is linear on the space spanned by the symbols $[W_i]$. The representations are as follows:
\begin{itemize}
\item
The ten first entry symbols $\{[W_i]\}_{i=1}^5\cup\{[W_j]\}_{j=16}^{20}$ are all related by permutations in ${\cal S}_5$. They decompose in the $1+4+5$ representations of ${\cal S}_5$. 
\item
The fifteen symbols $\{[W_i]\}_{i=6}^{15}\cup\{[W_j]\}_{j=21}^{25}$   are all related by permutations ${\cal S}_5$. They decompose in the $1+4+5+5'$ representations. 
\item
The five odd symbols $\{[W_i]\}_{i=26}^{30}$ transform in the irreducible $5'$ representation of ${\cal S}_5$.
\item
Finally, $[W_{31}]$ is invariant under ${\cal S}_5$. 
\end{itemize}

\subsection{Harmonic polylogarithms}
\label{app: harmonic polylogs}

The harmonic polylogarithms (HPL) form a very useful set of iterated integrals in one variable. We refer to \cite{Maitre:2005uu} as a convenient reference on the HPL functions. A HPL of weight $n$ in the variable $x$ is of the form $\HPL_{a_1,\ldots,a_n}(x)$, where $a_i\in \{-1,0,1\}$. They are defined iteratively by first setting
\beq
\begin{split}
&\HPL_1(x)=\int_{0}^xf_1(t)dt=-\int_{0}^xd\log(1-t)=-\log(1-x)\,,
\\ &\HPL_0(x)=\int^xf_0(t)dt=\int^x d\log(t)=\log(x)\,,
\\ &\HPL_{-1}(x)=\int_{0}^xf_{-1}(t)dt=\int_{0}^xd\log(1+t)=\log(1+x)\,,
\end{split}
\eeq
and then defining 
\beq
\HPL_{0^n}(x)\equiv \HPL_{\underbrace{0,\ldots, 0}_{n-\text{times}}}(x)=\frac{\log^n(x)}{n!}\,,\qquad \HPL_{a_1,\ldots, a_n}(x)=\int_{0}^xdtf_a(x')\HPL_{a_2,\ldots, a_n}(x')\,.
\eeq
We refer to \cite{Maitre:2005uu} for an introduction to the short-hand notation for the HPL functions $\HPL_{a_1,\ldots, a_n}$ with $|a_i|>1$. 

\subsection{The odd weight two pentagon functions} 
\label{AppOddFun}

In this appendix we provide functional basis of the nine dimensional subspace $\text{V}_{2,\text{odd}}$ of odd weight 2 non-planar pentagon functions, see table~\ref{tab:number of integrable symbols}. 
As we have already mentioned after \eqref{funep2}, by acting with permutations on the function \p{funep2} we obtain 30 different but linearly dependent  functions which cover $\text{V}_{2,\text{odd}}$.  However, we want to instead provide a smaller set of 10 functions which satisfy just one linear relation. For this, we use the following single-valued function
\begin{align}
\text{D}_2(z,\bar z) = \mathrm{Li}_2(z) - \mathrm{Li}_2(\bar z) + \frac{1}{2} \log( z\bar z)\left( \log(1-z) - \log(1-\bar z)\right)   
\end{align}
and use the shorthand notation $\text{D}_2(z) \equiv \text{D}_2(z,z^*)$.
We define the function $\text{F}_2$ via the linear combination 
\begin{align}
\text{F}_2(v_1,\ldots, v_5) = \text{D}_2(W_{26}) + \text{D}_2(W_{30}) - \text{D}_2(W_{26} W_{30})
\end{align}
We remind that in Minkowski kinematics we have $(W_j)^* = W_{j}^{-1}$ for $j=26,\ldots, 30$. The function $\text{F}_2$ has the same symbol (up to a factor $3$) as \p{funep2}. 
Acting by ${\cal S}_5$ permutations onto $\text{F}_2$ we obtain 10 different functions. They satisfy one linear relation which
is equivalent to
$$
\sum_{\sigma \in {\cal S}_5} \text{D}_2(\sigma(W_{26})) = 0\,,
$$
or more explicitly
\beq
\label{eq: 15 term identity for D2}
\sum_{j=1}^{5}\left[\text{D}_2\left(U_{j}\right) + \text{D}_2\left(\frac{1}{U_{j}U_{j+1}}\right)+ \text{D}_2\left(\frac{U_jU_{j+1}}{U_{j+3}}\right)\right]=0\,,
\eeq
where the in the above equation we have cyclically identified the odd letters, i.e. $U_{j}=W_{25+j}$ for $j=1,\ldots, 5$ with the relation $U_{j+5}\equiv U_j$. 

Polylogarithm identities similar to the 15-term one \eqref{eq: 15 term identity for D2} are intensely investigated in the literature \cite{Zagier:2007knq, LewinBook}. We do not know if \eqref{eq: 15 term identity for D2} is a new identity of if it follows from already known ones.

\providecommand{\href}[2]{#2}\begingroup\raggedright\endgroup
\end{document}